\documentclass[american,aps,pra,reprint,superscriptaddress,twocolumn]{revtex4-1}
\usepackage[unicode=true,pdfusetitle, bookmarks=true,bookmarksnumbered=false,bookmarksopen=false, breaklinks=false,pdfborder={0 0 0},backref=false,colorlinks=false] {hyperref}
\hypersetup{ colorlinks,linkcolor=myurlcolor,citecolor=myurlcolor,urlcolor=myurlcolor}
\usepackage{braket,colortbl,amsthm,amsmath,cleveref,amssymb,txfonts}
\definecolor{myurlcolor}{rgb}{0,0,0.7}
\usepackage{graphics,graphicx}
\usepackage{color}

\theoremstyle{plain}

\def\bea{\begin{eqnarray}}
\def\eea{\end{eqnarray}}
\def\ba{\begin{array}}
\def\ea{\end{array}}

\def\ket{\rangle}
\def\bra{\langle}
\def\beq{\begin{equation}}
\def\eeq{\end{equation}}

\begin{document}
\title{Non-Markovianity and entanglement detection}

\author{Sourav Chanduka}
\affiliation{Center for Computational Natural Science and Bio informatics,
International Institute of Information Technology, Gachibowli, Hyderabad, India.}
\author{Bihalan Bhattacharya} 
\email{bihalan@gmail.com}
\affiliation{S. N. Bose National Centre for Basic Sciences
Block - JD, Sector - III, Salt Lake City, Kolkata - 700 106}
\author{Rounak Mundra}
\affiliation{Center for Computational Natural Science and Bio informatics,
International Institute of Information Technology, Gachibowli, Hyderabad, India.} 
\author{Samyadeb Bhattacharya}
\email{samyadeb.b@iiit.ac.in}
\affiliation{Center for Security Theory and Algorithmic Research,
International Institute of Information Technology, Gachibowli, Hyderabad, India}
\author{Indranil Chakrabarty} 
\email{indranil.chakrabarty@iiit.ac.in}
\affiliation{Center for Security Theory and Algorithmic Research,
International Institute of Information Technology, Gachibowli, Hyderabad, India}


\begin{abstract}
\noindent We have established a novel method to detect non-Markovian indivisible quantum channels using structural physical approximation. We have shown that this method can be used to detect eternal non -Markovian operations. We have further established that harnessing eternal non-Markovianity, we can device a protocol to detect quantum entanglement.

\end{abstract}

\maketitle

\section{Introduction}
The theory of open quantum systems provides a powerful tool to study system-environment interactions, spawning decoherence, dissipation and other irreversible phenomena \citep{alicki,breuer}.
Inter and Intra qubit properties like entanglement, discord and coherence are also studied when one of the qubit undergoes decoherence in a system- bath interaction model \citep{chakrabarty2010study, bellomo, arend, samya1, samya2, nm1, nm3, nm4}.
 Recently, much efforts have been devoted to characterize quantum analogue of non-Markovian (NM) evolutions \citep{rivas1,breuerN,alonso,blp1,rhp1,bellomo,arend,samya1,samya2,samya3,wolf1,nm1,nm2,nm3,nm4,nm5,nm6}. It has been shown from various information theoretic and thermodynamic aspects, that NM can act as a powerful resource \citep{samya1,samya2,samya3}.
  Therefore, identification of non-Markovianity in a process is an extremely important area to study. However, it still remains an onerous task to construct a theory of distinguishing NM evolutions from its Markovian counterparts along with proposals of  experimentally feasible detection procedures. In most of the previous literature, non-Markovianity in terms of information backflow and divisibility breaking of a dynamical map is addressed by geometric or entropic distance based quantities \citep{rivas1,breuerN,alonso}. These measures, though offers a proper quantification, are not feasible measures in an experimental scenario. However, from the theory of entanglement \citep{wit6}, we know that hermitian operators are experimentally measurable and thus linear witnesses possess a much higher status from this perspective. Recently the present authors have constructed a convex resource theory of NM \citep{samyadeb}, creating that very opportunity of experimental verification, by exploring the geometry of NM dynamics in a similar manner of entanglement detection.\\

Detection of quantum entanglement is one of the most prevalent area of research in quantum information theory. It was Peres who introduced the partial transpose criterion (PT) for the detection of entanglement  and later Horodecki et.al 
proved that the criterion is  necessary and sufficient for $2 \otimes 2$ and $2 \otimes 3$ quantum system. However, the criterion is necessary for $d_1 \otimes d_2$ $(d_1, d_2 \geq 3)$ dimensions \citep{peres1996separability,horodecki2001separability}. The phenomena of witnessing entanglement \citep{wit2,wit3,wit4,wit5,wit7,wit8,wit9,wit10,JPCO21} is a stepping stone in the study of quantum information, offering versatile tools for experimental detection of entangled quantum states. Witnesses are hermitian operators and hence observables by construction, giving positive expectation value for all separable states; whereas negativity of the same signifies the existence of entanglement.  Not only detection of entanglement, witness operators were also used to detect various information processing tasks like teleportation and super dense coding \citep{ganguly2011entanglement,patro2017non,vempati2020witnessing}                                                               

In this article, we apply this methodology in open system dynamics, to construct NM witnesses from the similar footings of that of entanglement. 
Although, works have been done with the motivation of developing NM witnesses \citep{NMwit,witnm1,witnm2}, but in order to construct a proper NM detection theory, we need to have a convex and compact set of states beholding all Markovian operations. Though channel state duality \citep{jamil,choi} allows us to construct the set of states, due to the non-convex nature of divisible operations \citep{wolf1,wolf2}, constructing a theory of  linear witnesses is not possible in general.  We overcome this difficulty by ``small time interval" approximation, whence constructing the Choi states. This allows us to build a proper framework of linear witnesses for NM detection.

There are arguments on whether or not the indivisible operations exhaust all the non- Markovian operations \citep{modi2}. It is established that all Markovian operations are divisible, whereas the converse is not proven to be true. However, even if there exists NM operations which are CP-divisible, such operations can not generate resource back flow, and hence are not considered as resource \citep{samya4}. It is to mention that, in this paper, we will develop the theory of witneesses for NM operations which are indivisible. 
It is also very important here to mention that we are restricted to the set of operations having Lindblad type generators \citep{lindblad,gorini} of the form $\dot{\rho}(t)=\mathcal{L}_t(\rho(t))=\sum_{\alpha=1}^{n\leq d_S^2}\Gamma_{\alpha}(t)\left(L_{\alpha}\rho(t)L_{\alpha}^{\dagger}-\frac{1}{2}L_{\alpha}^{\dagger}L_{\alpha}\rho(t)-\frac{1}{2}\rho(t)L_{\alpha}^{\dagger}L_{\alpha}\right)$, with $\Gamma_{\alpha}(t)$s as the Lindblad coefficients, $A_{\alpha}$s as the Lindblad operators for a system with dimension $d_s$. For divisible operations we have $\Gamma_{\alpha}(t)\geq 0,~ \forall \alpha, t$ \citep{rhp1}. If the evolution is non-Markovian in the sense that it is indivisible, the $\Gamma_{\alpha}(t) < 0$, for some $\alpha$ at some instant of time $t$. According to reference \citep{rhp1}, to ensure complete positivity of the total dynamics, the condition $\int_0^T\Gamma_{\alpha}(t)dt\geq 0~ (\forall \alpha, T)$ must always hold. However, it has been shown later \citep{hall1} that this condition of positivity of all of the Lindblad coefficients over integration, can be relaxed for certain non-Markovian completely positive operations. For example, we can consider the following qubit depolarization operation 

\beq\label{hallA}
\frac{d\rho}{dt}=\sum_{i=1}^3\gamma_i\left(\sigma_i\rho\sigma_i-\rho\right),
\eeq
where $\sigma_i$ denotes the Pauli $\sigma$ matrices. If we now consider $\gamma_1(t)=\gamma_2(t)=1$ and $\gamma_3=-\tanh t$, then despite the fact that one of the Lindblad coefficient is negative throughout and hence violating the positivity condition upon integration, the concerning dephasing operation is still completely positive. This kind of operations with negative coefficients are coined as eternal non-Markovian operations. It has been shown that these eternal non-Markovian operations cannot be detected by either the divisibility based non-Markovianity measure or the information backflow based measure \citep{hall1}. In this article, we show that our method of non-Markovianity witness can successfully detect such eternal non-Markovian operations. We also find that eternal non-Markovian operations can give rise to positive but not completely positive maps and hence presents us with an opportunity to detect entanglement via the usage of non-Markovianity. 

The paper is constructed in  the following manner. In section II, we discuss the non-Markovianity witnessing and the application of structural physical approximation in that purpose. Then we discuss the detection of eternal non-Markovianity by our method. In section III, we then propose our method of entanglement detection by non-Markovianity and then finally we conclude in section IV with some possible implications. 

\section{Witnessing non-Markovianity and structural physical approximation}

In this section, we construct the process of witnessing non-Markovianity, based upon a previous formulation \citep{bihalanA} and further extending it. However, before going into the main results of this work, we first elucidate the properties of the set of Choi states for divisible operations. Then we develop the theory of linear NM witnessing. We further consider the geometry of the set of Choi states, to identify the possibility of generalized non-linear witnesses for NM detection. Then we conclude with stating the possible implications.\\

\noindent \textbf{Divisible operations and structure of Choi states:}Let us consider a quantum channel $\Lambda(t_2,t_1)$ which takes quantum state $\rho(t_1)$ at time $t_1$ to $\rho(t_2)$ at time $t_2$. Moreover let us consider that the channel has Lindblad type generator i.e. the  channel can be expressed as $\Lambda(t_2,t_1)\equiv \exp\left(\int_{t_1}^{t_2}\mathcal{L}_tdt\right)$.  Let $\mathbb{D}_{\mathcal{C}}$ represents the set of all such channels. It is well known that the quantum channels are in one to one correspondence with the corresponding Choi states. Exploiting this channel-state duality, we have an isomorphism between $\mathbb{D}_{\mathcal{C}}$ and the set of corresponding Choi states $\mathbb{C}_{\mathcal{A}}$. Now, given a map $\Lambda_{\mathcal{M}}(t_3,t_1)$, it is called divisible if we can find another linear map  $\Lambda_{\mathcal{M}}(t_3,t_2)$ on the state space such that 
\[ \Lambda_{\mathcal{M}}(t_3,t_1)=\Lambda_{\mathcal{M}}(t_3,t_2)\circ\Lambda_{\mathcal{M}}(t_2,t_1) \]
 with  $ t_3>t_2>t_1 \forall t_1, t_2, t_3$. Sometimes the map $\Lambda_{\mathcal{M}}(t_3,t_2)$ is called the propagator of the dynamics. A given map is called complete positive divisible or CP-divisible if the propagator map $\Lambda_{\mathcal{M}}(t_3,t_2)$ is also completely positive. In this case the Choi state $\mathcal{C}_{\mathcal{M}}(t_2,t_1)=\mathbb{I}\otimes\Lambda(t_2,t_1)|\phi\ket\bra\phi|$ is a valid density matrix for every instant of time, with $||\mathcal{C}_{\mathcal{M}}(t_2,t_1)||_1=1$, $\forall t_1,t_2$. Here $\mathbb{I}$ stands for the identity map acting on one subsystem and $|\phi\ket$ corresponds to a maximally entangled state in $d_s^2$ dimension and $||\cdot||_1=Tr[\sqrt{(\cdot)^{\dagger}(\cdot)}]$ is the trace norm. Breaking of CP-divisibility  of a dynamics is a signature of non-Markovian backflow of information \citep{rhp1,blp1}. Hence the CP divisible channels can be considered as memoryless Markovian channels. Therefore the set $\mathbb{F}_{\mathcal{M}}\subset\mathbb{C}_{\mathcal{A}}$ can be defined as $\mathbb{F}_{\mathcal{M}}=\{\mathcal{C}_{\mathcal{M}}(t_2, t_1)~|~~ ||\mathcal{C}_{\mathcal{M}}(t_2, t_1)||_1=1, \forall t_1, t_2\}$. This includes the Choi states for all CP-divisible operations. We recall the fact that $\mathbb{F}_{\mathcal{M}}$ is not convex in general. The convexity can be induced by considering $t_2=t_1+\epsilon$ and imposing the constraint of $\epsilon\rightarrow 0$  \citep{samyadeb}. The physical meaning of this approximation lies in observing the state in snapshots.  Note that  $\mathbb{F}_{\mathcal{M}}^{\epsilon}=\{\mathcal{C}_{\mathcal{M}}(t_1+\epsilon,t_1)~|~~ ||\mathcal{C}_{\mathcal{M}}(t_1+\epsilon, t_1)||_1=1, \forall t_1, \epsilon\}$ \\

\noindent \textbf{Non-Markovianity Witness:}  Let us consider the construction of NM witness using the techniques of entanglement theory. \\

\noindent \textbf{Definition:} A hermitian operator $W$ is said to be a NM witness if it satisfies following criteria: 
\begin{enumerate}
\item[1.] $Tr(W \mathcal{C}_{\mathcal{M}} )$ $\geqslant$ 0 $\forall$ $\mathcal{C}_{\mathcal{M}}$ $\in$
$\mathbb{F}_{\mathcal{M}}^\epsilon$.\\
\item[2.] There exists atleast one NM Choi state $\mathcal{C}_{\mathcal{N}}$ such that $Tr(W \mathcal{C}_{\mathcal{N}} )$ $<$ 0.
\end{enumerate}

It is clear from the definition, that a single witness can not detect all NMCS. The witness will depend on the NMCS, which one wishes to detect.\\

\noindent \textbf{Structural Physical Approximation:}
In this part, we formulate the protocol for experimental detection of information back flow.  The structural physical approximation (SPA)\citep{spa2,spa1} of a positive map is a convex mixture of a depolarizing map with the given map, so that the resulting map is complete positive. 
Consider a NCP map $\mathcal{M}$ acting on the Hilbert space $\mathcal{H}_d$ of dimension $d$. the following approximate map 
\[\mathcal{S}(\cdot)=\left[p\Theta+(1-p)\mathcal{M}\right](\cdot),\] with depolarizing map $\Theta(\rho)=\frac{\mathbb{I}_d}{d}$  in $\mathcal{H}_d$, can always be completely positive and hence experimentally implementable, over certain threshold value of $p$. Here $\mathbb{I}_d$ stands for identity operator on dimension d. An algorithm to find the optimal SPA map for a given positive map has  been prescribed in \citep{spa2}. 
Following the prescription , the optimal SPA map corresponding to the map $\mathcal{M}$ is given by
\beq\label{sp1}
\mathcal{S}^{opt} = p^* \Theta + (1-p^*) \mathcal{M'} 
\eeq
where $p^* = \frac{\lambda d d^{'} \beta_{\Lambda_{\alpha}}^{-1}}{\lambda d d^{'} \beta_{\mathcal{M}}^{-1}+ 1}$, $\lambda$ is the minimum eigenvalue of  $\chi=[\mathbb{I} \otimes\mathcal{M}](|\phi_d\ket\bra\phi_d|)$ , and $\mathcal{M'}= \beta_{\mathcal{M}}^{-1} \mathcal{M}$  is the re-scaling of the original map. Here $d$  and $d'$ are the input and output dimension of the map $\mathcal{M}$, $\vert \phi_d \ket$ denotes the maximally entangled state in dimension d and $\beta_{\mathcal{M}}$ is a rescaling parameter. Assuming  the trace preservation property of the map, the value of $ \beta_{\mathcal{M}} = 1$.\\\\

\noindent \textbf{Information backflow witness for qubit systems:-} In the following, we construct information backflow witnesses for two dimensional systems. We also give some particular examples for specific qubit dynamical maps. But before going into further study, we prove the following proposition, which will be utilized frequently, in the later study.\\

\noindent\textbf{Proposition 1:} For any arbitrary dimensional system, the following identity is always true when $G$ is Harmitian.
\[Tr\left[|\alpha\ket\bra\alpha |\mathbb{I}\otimes\mathcal{N}(\rho)\right]=Tr\left[\mathbb{I}\otimes\mathcal{N}(\vert\alpha\ket\bra\alpha \vert)\rho\right],\] where $|\alpha\ket$ is any arbitrary pure state in $d\time 3$ dimension and
\[
\begin{array}{ll}
\mathcal{N}(\rho)=
\rho
+\\
\Gamma(t)\left[\left(G\rho G^{\dagger}
-\frac{1}{2}(G^{\dagger}G\rho +\rho G^{\dagger}G)\right)\right].
\end{array}
\]

\proof Consider the operation \[\mathcal{R}(\rho)=\rho+\left[\Gamma(t)\left(\mathcal{A}\rho\mathcal{A}^{\dagger}-\frac{1}{2}(\mathcal{A}^{\dagger}\mathcal{A}\rho+\rho\mathcal{A}^{\dagger}\mathcal{A}\right)\right],\] with $\mathcal{A}$ being any Hermitian operator on the Hilbert space $\mathcal{H}_d\otimes\mathcal{H}_d$. Now using the property $Tr[AB]=Tr[BA]$ for any two matrices $A$ and $B$ and $Tr[\rho]=Tr[|\alpha\ket\bra\alpha |]=1$, we show that
\[
Tr[|\alpha\ket\bra\alpha |\mathcal{R}(\rho)]=Tr[\rho\mathcal{R}(|\alpha\ket\bra\alpha |)].
\]
Since $\mathbb{I}\otimes\mathcal{N}(\rho)$ falls into the category of $\mathcal{R}(\rho)$ with $\mathcal{A}=\mathbb{I}_d\otimes G$, this proves the proposition. \qed

\noindent Therefore for the case of unital operations with Harmitian Lindblad operators, we can device our method to detect information backflow witness.\\

\noindent \textit{Construction of witness:-} We now move on to construct the witness operator  for detecting non-Makovianity, on the backdrop of SPA protocol. Let us consider the maximally entangled state $\sigma$. Therefore the operator $\tilde{\sigma}=\mathcal{S}^{\mathcal{N}}(\sigma)$ is always a positive semi definite matrix, possessing all the properties of a quantum state. The minimum eigenvalue $\lambda_{min}$ determines, whether the original map is indivisible or not. If the minimum eigenvalue $\lambda_{min} < \omega$, the operation is indivisible. The threshold value $\omega$ for a given dimensional system is to be determined by the protocol presented in Ref. \citep{spa2}. We now define an operator $\mathcal{O}=\tilde{\sigma}-\nu\mathbb{I}\otimes\mathcal{N}(\sigma)$, where $\nu$ is a suitable chosen real number. Let us now consider $|\tau\ket$, to be the eigenvector corresponding to the minimum eigenvalue of $\tilde{\sigma}$. We choose $\nu$ such that \[\bra\tau |\mathcal{O}|\tau\ket=Tr[(\tilde{\sigma}-\nu\mathbb{I}\otimes\mathcal{N}(\sigma))|\tau\ket\bra\tau |]=\omega.\]

Note that a hermitian operator $\mathcal{W}$ is a NM witness if it produces non negative expectation on all Markovian Choi states and there exists atleast one NM Choi matrix on which $\mathcal{W}$ gives negative expectation. Considering the hermitian operator $\mathcal{W}=\nu\mathbb{I}\otimes\mathcal{N}(|\tau\ket\bra\tau |)$ and

Using \textbf{Proposition 1}, we get 
\[
Tr[\mathcal{W}\sigma]=Tr[\nu\mathbb{I}\otimes\mathcal{N}(|\tau\ket\bra\tau |)\sigma]=Tr[\tilde{\sigma}|\tau\ket\bra\tau |]-\omega.
\]
If we consider $\mu_{min}$ to be the minimum eigenvalue of $\tilde{\sigma}$, the we have $Tr[\mathcal{W}\sigma]=\mu_{min}-\omega$. It is then straight forward to check that $Tr[\mathcal{W}_{\mathcal{M}}\sigma]\geq 0$, for all divisible operations $\mathcal{W}_{\mathcal{M}}$ and $Tr[\mathcal{W}_{\mathcal{N}}\sigma] < 0$ for indivisible operations $\mathcal{W}_{\mathcal{N}}$. Therefore, 
\beq\label{wit1}
\mathcal{W}=\nu\mathbb{I}\otimes\mathcal{N}(|\tau\ket\bra\tau |),
\eeq
acts as a witness operator, detecting whether the operation under consideration is NM or not. Since $\omega$ is the threshold value for positivity, the witness $\mathcal{W}$ is optimal. Following Ref.\citep{spa2}, we now describe the protocol to find the threshold value $\omega$. Consider the SPA map $\mathcal{S}^{\mathcal{N}}(\sigma)$ of the entangled state $\sigma$.The error term can be written as 
\[\Delta(\sigma)=\mathcal{S}^{\mathcal{N}}(\sigma)-\gamma_{\sigma}\mathbb{I}\otimes\mathcal{N}(\sigma),\]
where the error term satisfies the invariance condition $\Delta(\sigma)=\delta_{\sigma}\mathbb{I}$. Here $\gamma_{\sigma}$ and $\delta_{\sigma}$ are in general state dependent quantity, whose optimal value gives us $\nu$ and $\omega$ respectively. Therefore the SPA map can be be written as 
\[\mathcal{S}^{\mathcal{N}}(\sigma)=\delta_{\sigma}\mathbb{I}+\gamma_{\sigma}\mathbb{I}\otimes\mathcal{N}(\sigma),\]
with the functions $\gamma_{\sigma},~\delta_{\sigma}\geq 0$. Now following the protocol described in Ref. \citep{spa2}, we find that the optimal decomposition can be written as 
\beq\label{opt1}
\mathcal{S}^{\mathcal{N}}_{opt}(\sigma)= \omega\mathbb{I}+\nu\mathbb{I}\otimes\mathcal{N}(\sigma),
\eeq
with 
\[\nu=\frac{1}{|\lambda_-| d^2+1},~~\omega=\frac{|\lambda_-| d^2}{|\lambda_-| d^2+1}.\]

Let us now illustrate the above mentioned procedure with the example of qubit dephasing dynamics.\\\\

\noindent \textit{Qubit dephasing channels :-} Let us first consider qubit dephasing channels having Lindblad type generators of the form
\[\mathcal{L}_D(\cdot)=\Gamma_D(t)\left( \sigma_z(\cdot)\sigma_z-(\cdot) \right),\]
with $\sigma_i~~(i=1,2,3)$ being the Pauli matrices and $\Gamma_D(t)$ is Lindblad coefficient. This generator will correspond to an indivisible dephasing operation, if the dephasing parameter has got some negative regions. Therefore under the small time interval approximation, the map corresponding to this dephasing operation will be $\mathcal{N}_D(\cdot)=(\mathbb{I}+\epsilon\mathcal{L}_D)(\cdot)$. The negative eigenvalue will be equal to $\Delta=-\epsilon|\Gamma_D(t)|$, corresponding to the eigenvector $|\lambda_D^-\ket=(-1,0,0,1)^T$.

For this map, if we apply the SPA map as given in \eqref{sp1}, we find that the optimal value of mixing parameter is 
\beq\label{SPA1a}
p_{min}^D= \frac{4\epsilon|\Gamma_D(t)|}{1+4\epsilon|\Gamma_D(t)|}.
\eeq

Let us now construct the witness operator corresponding to this evolution. We consider the SPA map $S^{\mathcal{N}}_D(\rho)$ with $p>p_{min}^D$. In accordance with the theory of entanglement, we consider 
\[\omega= \frac{4\epsilon|\Gamma_D(t)|}{1+4\epsilon|\Gamma_D(t)|} ~~~~~ \nu= \frac{1}{1+4\epsilon|\Gamma_D(t)|}\] 
Since the protocol we have described is valid for any state, we consider $\sigma=|\lambda_D^-\ket\bra\lambda_D^-|$.\\\\

\section{Eternal non-Markovianity and entanglement detection}

 In this section we shall discuss an important class of non-Markovian operations and its connection with entanglement detection.
 Markovianity is closely connected with complete positive divisibility. CP indivisibility implies non-Markovianity in general. But as mentioned in the introduction section, there are examples of dynamics which is completely positive through out the whole time interval but nature of which is non-markovian. Such non-Markovian operations are known as eternal non-Markovian operations. Such non-Markovianity can not be detected by trace distance measure or Bures distance based measure or entanglement measure of non-Markovianity. in this section we shall illustrate eternal non-Markovianity and its connection with entanglement detection using example.\\\\ 
Let us consider the operation with the Lindblad equation:
\[\Lambda_{dep}(\rho)= \gamma_x (\sigma_x \rho \sigma_x-\rho)+\gamma_y (\sigma_y \rho \sigma_y-\rho)+\gamma_z (\sigma_z \rho \sigma_z-\rho)\]

This is actually a qubit depolarising operation. Now consider $\gamma_x=\gamma_y=1$ and $\gamma_z=-\tanh t$.\\
To witness this non-Markovianity, we use a witness operator, $\mathcal{W}_z= \vert \chi \ket \bra \chi \vert$ where $\chi= (-1,0,0,1)^T$. This type of non-Markovianity can not be detected by Rivas-Huelga-Plenio (RHP) measure \citep{rhp1}. We would like to construct the SPA of the corresponding map. Following the prescription of \citep{spa2} we calculate the value of $\omega$ and $\nu$ and we find that
\[\omega=\frac{4 \epsilon \tanh t}{1+4 \epsilon \tanh t} ~~~~~~~~~~~~~ \nu= \frac{1}{1+4 \epsilon \tanh t} \]\\

\noindent \textbf{Entanglement Detection:} 
It is well known that a positive but not completely positive maps play an important role in detecting entanglement. A quantum state $\sigma$ is said to be separable if and only if for every positive map $\Lambda$, the operator $\mathbb{I} \otimes \Lambda (\sigma)$ is a positive semidefinite. Certainly this criterion for detecting entanglement is extremely hard in practice. As mentioned in the introduction for two qubit states and qubit qutrit states transposition map plays an universal role to detect entanglement. But from two qutrit system onward existence of PPT entanglement which is considered to be an weak form of entanglement is very hard to detect. Transposition map can not detect such entanglement.  Efforts from both mathematics and physics community \citep{Stinespring55, arveson69, Choi75, Stormer82, Worono76, Cho92, Ha03, Majewski01, Sarbicki12, Miller15, JPCO21, Chruscinski12} have been able to shade light on the theory of positive maps and high dimensional entanglement. Still the the study of positive maps and entanglement detection is an active area of research as the structure of positive maps is not all clear even in low dimensions. Here we detect entanglement in a two qubit state using a family of positive maps. It is true that the same could have been done using transposition map but the main reason to present the example of the map is to show the connection between non-Markovianity and entanglement detection. It is well known that the Lindblad form of a map plays an important role in non-Markovianity.   Let us consider a general two parameter family of maps motivated from the Lindblad structure,
\[ \Lambda (\rho)= (\mathbb{I}+ \mathcal{L})(\rho)\] where 
\[\mathcal{L} (\rho)= \gamma_1 (\sigma_x \rho \sigma_x-\rho)+\gamma_1 (\sigma_y \rho \sigma_y-\rho)+\gamma_2 (\sigma_z \rho \sigma_z-\rho)\]

It can be shown that the above map is positive for $0 \leqslant \gamma_1 \leqslant \frac{1}{2}$ and $0 \leqslant \gamma_2 < 1$ and not completely positive whenever $\gamma_2 \geqslant 0$. Hence it can be used to detect entanglement. It is well known that the Werner class of states given by
\[\mathbb{W}_p = p \vert \psi^-\ket \bra \psi^-\vert +  (1-p)\frac{\mathbb{I}_2\otimes \mathbb{I}_2 }{4} \] 
is entangled for $p > \frac{1}{3}$. We have found that for the special case of the above mentioned positive map if $\gamma_1=\frac{1}{2} $ and $\gamma_2=\frac{1}{2} $, it can detect the full range of entangled state in the Werner class. Moreover it is interesting to note that even if we consider general $\gamma_1$ and $\gamma_2$  in the range where the above map is positive but not completely positive; it can detect the entanglement for the full range of the entangled Werner class. 

\section{Conclusion}

In this work, we have first constructed a novel method to witness non-Markovian quantum channels which are indivisible, by using a the SPA protocol. We have further observed that a special type of non-Markovian channels namely the eternal non-Markovian channels generate non complete positive maps within its dynamical evolution and hence can be utilized to detect entangled states. This gives us a unique avenue of research which may lead us to dynamical detection of entanglement in much more complex quantum systems.

\bibliographystyle{apsrev4-1}
\bibliography{sister1}

\end{document}